\documentclass[aps,10pt,a4paper,superscriptaddress,twocolumn,showpacs,showkeys,notitlepage]{revtex4}

\usepackage{amsfonts}
\usepackage{amssymb}
\usepackage[dvips]{graphicx}
\usepackage{hyperref}
\usepackage{amsmath}

\begin{document}

\title{Statistical properties of the truncated state with random coefficients%
}
\author{W. B. Cardoso}
\email{wesleybcardoso@gmail.com}
\affiliation{Instituto de F\'{\i}sica, Universidade Federal de Goi\'{a}s, CEP 74.001-970,
Goi\^{a}nia (GO), Brazil.}
\author{N. G. de Almeida}
\affiliation{Instituto de F\'{\i}sica, Universidade Federal de Goi\'{a}s, CEP 74.001-970,
Goi\^{a}nia (GO), Brazil.}
\affiliation{N\'{u}cleo de Pesquisas em F\'{\i}sica, Universidade Cat\'{o}lica de Goi\'{a}%
s, CEP 74.605-220, Goi\^{a}nia (GO), Brazil.}

\begin{abstract}
In this paper we introduce the truncated state with random coefficients
(TSRC). As the coefficients of the TSRC have in principle no algorithm to
produce them, our question is concerned about to what type of properties
will characterize the TSRC. A general method to engineer TSRC in the
running-wave domain is employed, which includes the errors due to the
nonidealities of detectors and photocounts.
\end{abstract}

\maketitle

\section{Introduction}

The generation of new states of the light field turned out to be an
important topic in quantum optics in the last years, such as quantum
teleportation \cite{bennett93}, quantum computation \cite{Kane98}, quantum
communication \cite{Pellizzari97}, quantum cryptography \cite{Gisin02},
quantum lithography \cite{Bjork01}, decoherence of states \cite{Zurek91},
etc. The usefulness and relevance of quantum states, to give a few examples,
cover the study of i) quantum decoherence effects in mesoscopic fields \cite%
{harochegato}; ii) entangled states and quantum correlations \cite{brune};
interference in the phase space \cite{bennett2}; collapses and revivals of
the atomic inversion \cite{narozhny}; engineering of (quantum states)
reservoir \cite{zoller}; etc. Also, it is worth mentioning the importance of
the statistical properties of one state to determine some relevant
properties of another \cite{barnett} as well as using specific quantum
states as input to engineer a desired state \cite{serra}.

Here we introduce a truncated state having random coefficients (TSRC). This
state is characterized by a constant phase relation between the relative
Fock states, but with coefficients describing probability amplitudes
obtained through some random number generator. Features of this state are
studied by analyzing several of its statistical properties. We believe the
TSRC can be a useful tool to test sequence of numbers whose generation is
unknown from the observer. Indeed, preliminary results have corroborated
this expectation \cite{wesley}.

This paper is organized as follows: in Section II we introduce the TSRC and
Section III we analyze the behaviour of some statistical properties of the
TSRC when the Hilbert space is increased. In Section IV we show how to
engineer TSRC in the running-wave domain, and Section V is devoted to the
study of losses due to nonidealities of the detectors in the whole process
of engineering. Finally, in Section VI we present our conclusions.

\section{Truncated states with random coefficients (TSRC)}

We define the TSRC as
\begin{equation}
\left\vert TSRC\right\rangle =\mathcal{N}\sum\limits_{n=0}^{N}r_{n}e^{in%
\theta }\left\vert n\right\rangle
\end{equation}%
where $\mathcal{N}$ is a normalization constant and $C_{n}=r_{n}e^{in\theta
} $ is a random complex coefficient (polar form), given by some random
number generator (RNG). It is worthwhile to note that TSRC differs from
(mixed) thermal states (MTS) and Pure States Having Thermal Photon
Distribution (PSTD)\cite{Marhic}, since these states have an algorithm for
the coefficients $C_{n}{\prime }s$. In addition, although TSRC is a pure
state, its statistical properties are very different from the PSTD: for
example, the TSRC does not exhibit squeezing for $N\gtrsim 12$, as shown in
the next section.

\section{Statistical properties of TSRC}

\subsection{Photon Number Distribution}

Since the expansion of TSRC is known in the number state $\left\vert
n\right\rangle $, we have
\begin{equation}
P_{n}=\mathcal{N}^{2}\left\vert C_{n}\right\vert ^{2}.
\end{equation}%
Fig. $1-3$ show the plots of the photon-number distribution, $P_{n}$ versus $%
n$, for TSRC. Although the generation of \ the TSRC be virtually impossible
for Hilbert space dimensions $10^{4}$, due to nonidealities of beam
splitters and detectors causing decoherence effects as well as due to low
success probability (see section IV), it can be useful to study the behavior
of TSRC for large $N$, specially when we are interested in investigate the
properties of sequence of numbers \cite{wesley}. Fig$-3$, which is a
expanded view of the Fig. $2$, shows the same behavior of that of the Fig.$1$%
. As expected, when $N$ is large enough (Fig. 2 ), the difference from one
peak to another becomes negligible. For small values of $n$, running the RNG
several times will give different sets of numbers with different ${P_{n}}%
^{\prime }s$. Therefore, it is important to keep in mind that our proposal
to generate TSRC is based on a particular set of numbers obtained by a
single run of the RNG. Thus, these numbers, having no algorithm producing
them, a question emerging is: what type of properties will characterize a
state having a random probability amplitude? These properties will be
investigated in the next subsections.

\begin{figure}[tb]
\includegraphics[height=7.5cm, width=7.5cm]{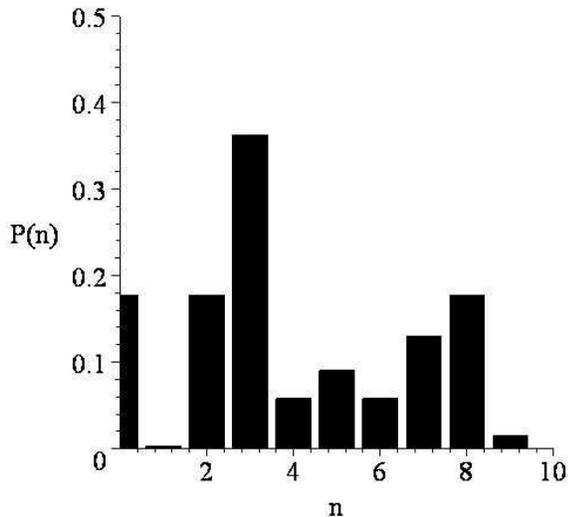}
\caption{The photon number distribution $P_{n}$ of TSRC versus n for the
dimension of Hilbert space $N=10$. This figure was obtained by a single run
of the RNG.}
\end{figure}

\begin{figure}[tb]
\includegraphics[{height=7.5cm, width=7.5cm}]{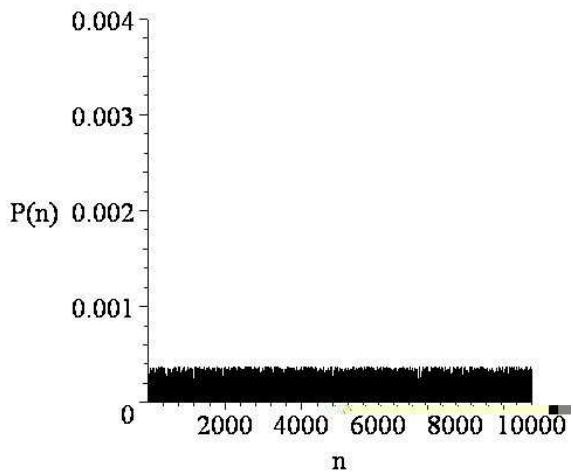}
\caption{The photon number distribution $P_{n}$ of TSRC versus $n$. As
expected, when $N$ is large the difference from one peak to another becomes
negligible. The dimension of Hilbert space here is $N=10000$. This figure
was obtained by a single run of the RNG.}
\end{figure}

\begin{figure}[tb]
\includegraphics[{height=7.5cm, width=7.5cm}]{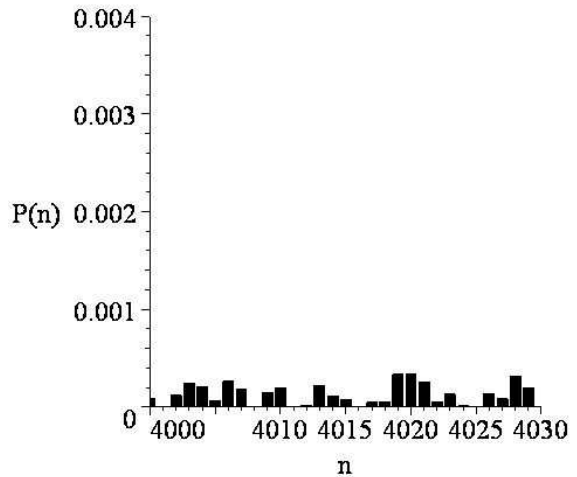}
\caption{Expanded view of the photon number distribution $P_{n}$ of \ the
TSRC versus $n$ in the Fig.$2$. This figure was obtained by a single run of
the RNG.}
\end{figure}

\subsection{Average number and variance}

The average number $\left\langle \hat{n}\right\rangle $ as well as the
variance in TSRC is obtained directly from
\begin{equation}
\left\langle \hat{n}\right\rangle =\sum_{n=0}^{N}P(n)n\text{,}
\end{equation}%
and

\begin{equation}
\left\langle \Delta \hat{n}\right\rangle =\sqrt{\left\langle \hat{n}%
^{2}\right\rangle -\left\langle \hat{n}\right\rangle ^{2}}\text{.}
\end{equation}

Fig. $4$, $5$ show the plots of $\left\langle \hat{n}\right\rangle $ and $%
\left\langle \Delta \hat{n}\right\rangle $ as a function of the Hilbert
space dimension $N$. Note from Fig.$3$ that $\left\langle \hat{n}%
\right\rangle $ increases irregularly with $N$ when a single realization is
considered, but tends to increase linearly with $N$ when several
realizations are considered. As can be seen from Fig.$5$, the increasing of $%
\left\langle \Delta \hat{n}\right\rangle $ with $N$ is irregular,
irrespective of the number of realizations.

\begin{figure}[tb]
\includegraphics[{height=6.5cm, width=8cm}]{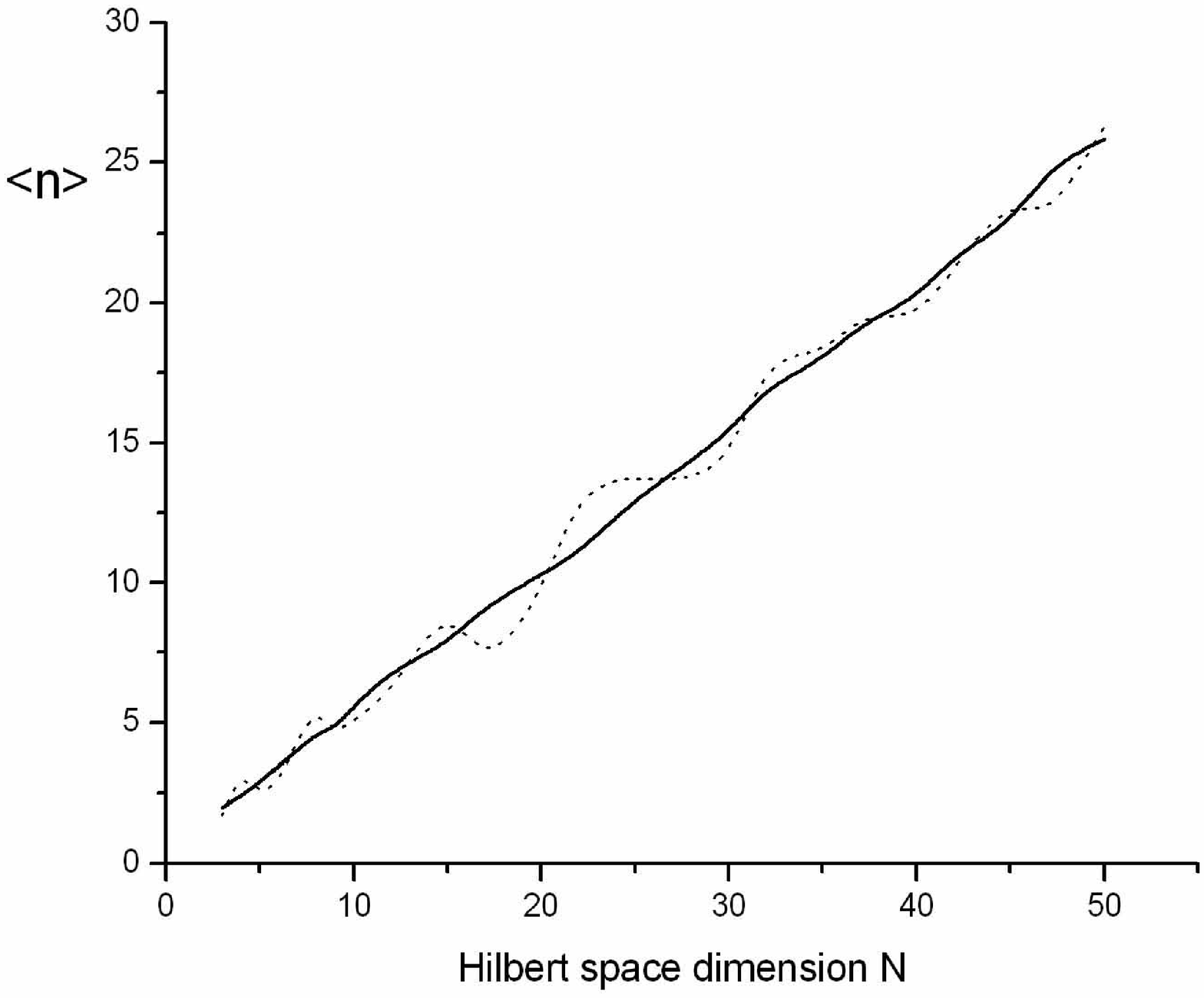}
\caption{The average number $<n>$ versus the dimension $N$ of the Hilbert
space. Solid line was obtained from 30 realizations of the RNG. Dots refer
to a single one.}
\end{figure}

\begin{figure}[tb]
\includegraphics[{height=6.5cm, width=8cm}]{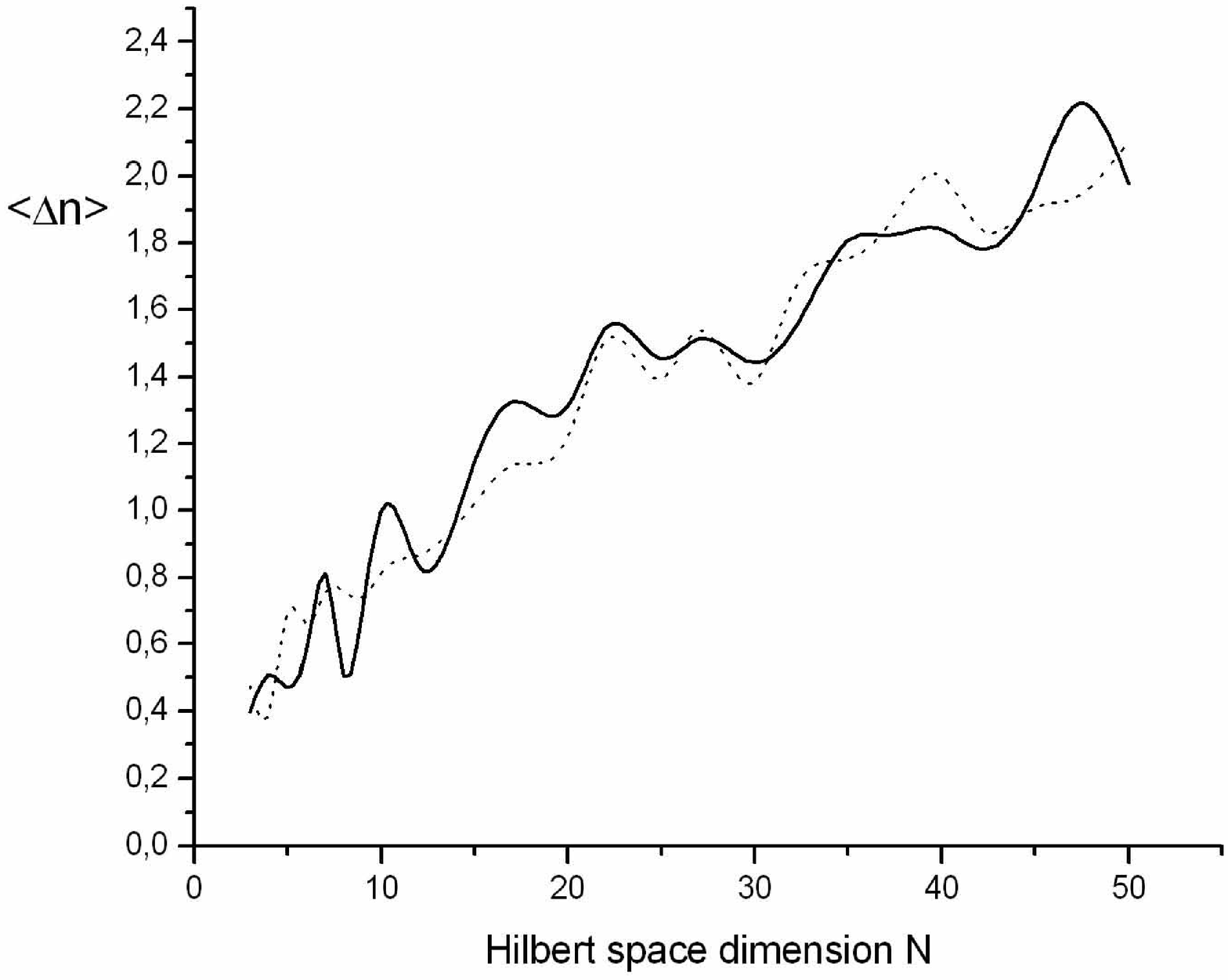}
\caption{The variance $\langle \Delta n \rangle$ versus the dimension $N$ of
the Hilbert space.}
\end{figure}

\subsection{Mandel parameter and second order correlation function}

Mandel Q parameter is defined as

\begin{equation}
Q=\frac{\left( \Delta \hat{n}^{2}-\left\langle \hat{n}\right\rangle \right)
}{\left\langle \hat{n}\right\rangle }\text{,}
\end{equation}%
while the second order correlation function $g^{\left( 2\right) }(0)$ is
\begin{equation}
g^{\left( 2\right) }(0)=\frac{\left( \left\langle \hat{n}^{2}\right\rangle
-\left\langle \hat{n}\right\rangle \right) }{\left\langle \hat{n}%
\right\rangle ^{2}}\text{,}
\end{equation}%
and for $Q<0$ ($Q>0$) the state is said to be sub-poissonian
(super-poissonian). Also, the $Q$ parameter and the second order correlation
function $g^{\left( 2\right) }$ are related by \cite{walls}
\begin{equation}
Q=\left[ g^{\left( 2\right) }(0)-1\right] \left\langle \hat{n}\right\rangle
\text{.}  \label{qg}
\end{equation}%
It is well known that if $g^{\left( 2\right) }(0)<1$\ then the
Glauber-Sudarshan P-function assumes negative values, thus differing from
usual probability distribution function. Beside that, by Eq. (\ref{qg}) is
readily seen that $g^{\left( 2\right) }(0)<1$ implies $Q<0$. As for a
coherent state $Q=0$, a state is said to be a \textquotedblleft
classical\textquotedblright\ one if $Q>0$.

Fig. $6$ and $7$ show the plots of the $Q$ parameter and the correlation
function $g^{\left( 2\right) }(0)$ versus $N$. It is interesting to note
that the TSRC is predominantly super-poissonian ($Q>0$ and $g^{\left(
2\right) }(0\dot{)}>1$) , then being a \textquotedblleft
classical\textquotedblright\ state in this aspect, for $N\gtrsim 12$.
Nevertheless, for small values of \ $N$ ($N<12$), the $Q$ parameter presents
significant occurrence of values less than $0$, showing sub-poissonian
statistics and thus being associated with a \textquotedblleft quantum
state\textquotedblright . However, when $N$ becomes larger than $12$, TSRC\
becomes super-poissonian. Also, note that for a single realization of the
RNG these both functions present an oscillatory behavior, while, for several
realizations, Fig.$6$ (solid line) shows an interesting linear dependence
between $Q$ and $N$. This will be further investigated in the following
subsection. On the other hand, Fig.$7$ indicates an asymptotic behavior of \
$g^{\left( 2\right) }(0\dot{)}$ when $N$ is increased. For $N=10^{4}$, we
find $g^{\left( 2\right) }(0\dot{)}=1.332585322$.

\begin{figure}[tb]
\includegraphics[{height=6.5cm, width=8cm}]{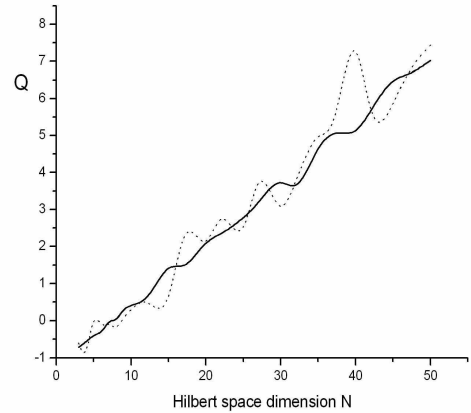}
\caption{The $Q$ parameter versus $N$. For $N\gtrsim 12$, $Q>0$ and there is
a linear dependence between $Q$ and $N$. Solid line was obtained from 30
realizations of the RNG. Dots refer to a single one.}
\end{figure}

\begin{figure}[tb]
\includegraphics[{height=6.5cm, width=8cm}]{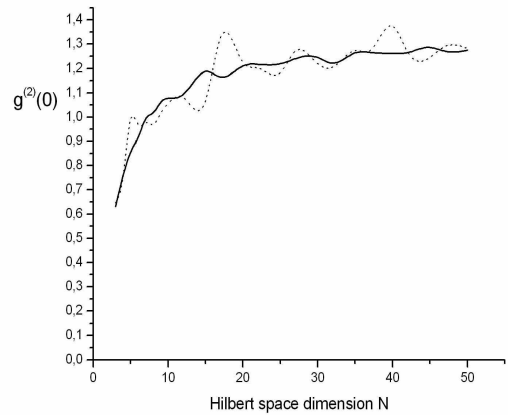}
\caption{The $g^{(2)}(0)$ correlation function versus $N$. Solid line was
obtained from 30 realizations of the RNG. Dots refer to a single one.}
\end{figure}

\subsection{Quadrature and variance}

Quadrature operators are defined as
\begin{eqnarray}
X_{1} &=&\frac{1}{2}\left( a+a^{\dagger }\right) ; \\
X_{2} &=&\frac{1}{2i}\left( a-a^{\dagger }\right) ,
\end{eqnarray}%
where $a$ ($a^{\dagger }$) \ is the annihilation (creation) operator in Fock
space. Quantum effects arise when the variance of one of the two quadrature
attains a value $\Delta X_{i}<0.5$, $i=1,2$. Fig. $8$ and $9$ show the plots
of quadrature variance $\Delta X_{i}=\sqrt{\left\langle
X_{i}^{2}\right\rangle -\left\langle X_{i}\right\rangle ^{2}}$ versus $N$.
Note the same oscillatory behavior when a single realization of RNG is
considered. For small $N$ ($N<12$),\ TSRC exhibits squeezing with a
significant frequency (running RNG several times). On the other hand, when $%
N $ increases, squeezing becomes a rare event, and can be set a null event
for $N\gtrsim 12$, coinciding with the departure from sub-poissonian
statistic discussed above.

\begin{figure}[tb]
\includegraphics[{height=6.5cm, width=8cm}]{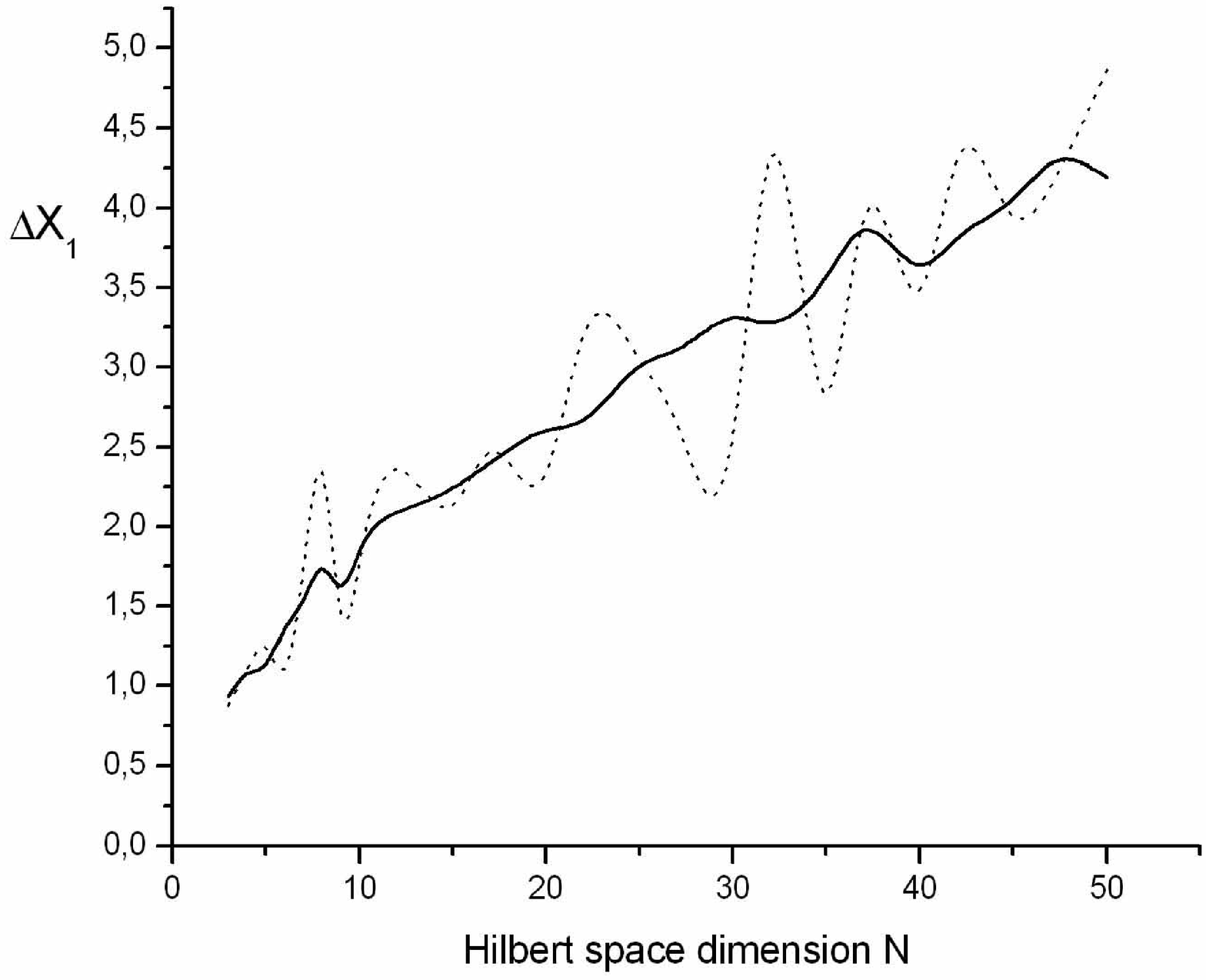}
\caption{The $\Delta X_{1}$ quadrature variance versus the dimension $N$ of
the Hilbert space. For $N\gtrsim 12$, squeezing can be considered a null
event. Solid line was obtained from 30 realizations of the RNG. Dots refer
to a single one.}
\end{figure}

\begin{figure}[tb]
\includegraphics[{height=6.5cm, width=8cm}]{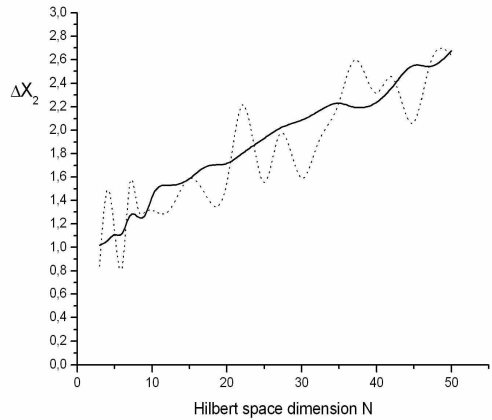}
\caption{The $\Delta X_{2}$ quadrature variance versus the dimension $N$ of
the Hilbert space. For $N\gtrsim 12$, squeezing can be considered a null
event. Solid line was obtained from 30 realizations of the RNG. Dots refer
to a single one.}
\end{figure}

\subsection{Entropy}

Fig. $10$ shows the plots of Shannon entropy $S=\sum
\limits_{n=0}^{N-1}P_{n}\ln P_{n}$ versus $N$. It shows a steady
increasing
of \ $S$. In fact, increasing $N$ arbitrarily, we have found no bound for $S$%
, thus indicating no maximum for Shannon entropy.

\begin{figure}[tb]
\includegraphics[{height=6.5cm, width=8cm}]{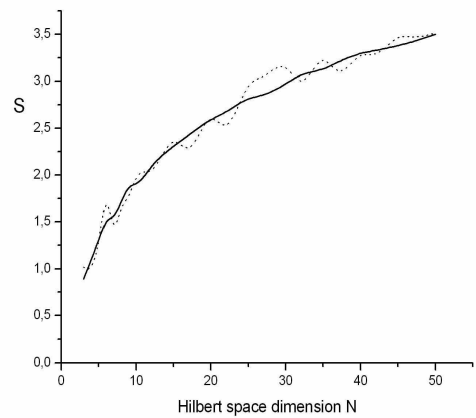}
\caption{The $S$ Shannon entropy versus the dimension $N$ of the Hilbert
space. Solid line was obtained from 30 realizations of the RNG. Dots refer
to a single one.}
\end{figure}

\subsection{Husimi Q function}

The Husimi Q-function for TSRC is given by
\begin{equation}
Q_{\left\vert TSRC\right\rangle }(\beta )=\frac{1}{\pi }\left\vert
\left\langle \beta |\Psi \right\rangle \right\vert ^{2}\text{.}
\end{equation}%
Fig.$11$ shows the Husimi Q-function for $N=15$. It is to be noted that this
function will change each time we run the RNG even for a fixed $N$. Thus, as
remarked above, we refer to a particular fixed set of numbers assumed as
having no law generating them.

\begin{figure}[tb]
\includegraphics[{height=8cm, width=8cm}]{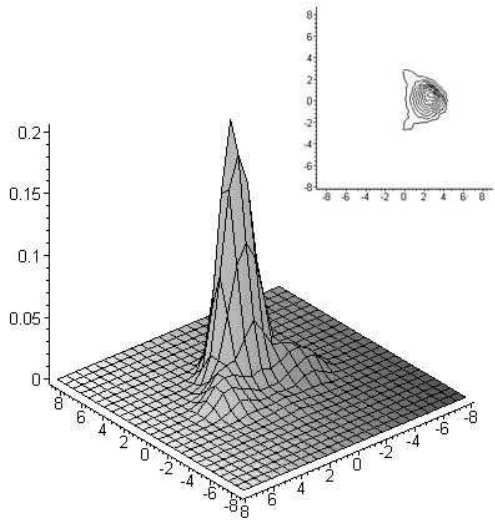}
\caption{Husimi Q-function for the TSRC taking the dimension of the Hilbert
space $N=15$.}
\end{figure}

\section{Generation of TSRC}

Schemes for generation of an arbitrary state can be found in various
contexts, as for example trapped ions \cite{serra2} and cavity QED \cite%
{serra,vogel}. However, due to the severe limitation imposed by coherence
loss and damping, we will employ the scheme introduced by Dakna et al. \cite%
{dakna} in the realm of running wave field. For brevity, the present
application only shows the relevant steps of Ref.\cite{dakna}, where the
reader will find more details. In this scheme, a desired state $\left\vert
\Psi \right\rangle $ composed of a finite number of Fock states $\left\vert
n\right\rangle $ can be written as%
\begin{eqnarray}
\left\vert \Psi \right\rangle &=&\sum_{n=0}^{N}C_{n}\left\vert
n\right\rangle =\frac{C_{N}}{\sqrt{N!}}\prod_{n=1}^{N}\left( \hat{a}%
^{+}-\beta _{n}^{\ast }\right) \left\vert 0\right\rangle  \notag \\
&=&\frac{C_{N}}{\sqrt{N!}}\prod_{k=1}^{N}\hat{D}(\beta _{k})\hat{a}^{+}\hat{D%
}(\beta _{k})\left\vert 0\right\rangle ,  \label{teorica}
\end{eqnarray}%
where $\hat{D}(\beta _{n})$ stands for the displacement operator and the $%
\beta _{n}$ are the roots of the polynomial equation
\begin{equation}
\sum\limits_{n=0}^{N}C_{n}\beta ^{n}=0.  \label{raizes}
\end{equation}

According to the experimental setup shown in the Fig.1 of Ref.\cite{dakna},
we have (assuming $0$-photon registered in all detectors) the outcome%
\begin{equation}
\left\vert \Psi \right\rangle \sim \prod_{k=1}^{N}D(\alpha _{k+1})\hat{a}%
^{+}T^{\hat{n}}D(\alpha _{k})\text{ }\left\vert 0\right\rangle ,
\label{experimental}
\end{equation}%
where $T$ is the transmittance of the beam splitter and ${\alpha }_{k}$ are
experimental parameters. After some algebra, the Eq.(\ref{teorica}) and Eq.(%
\ref{experimental}) can be connected. In this way, one shows that they
become identical when ${\alpha }_{1}=-\sum_{l=1}^{N}T^{-l}{\alpha }_{l+1}$
and $\alpha _{k}=T^{\ast N-k+1}(\beta _{k-1}-\beta _{k})$ for $k=2,3,4\ldots
N$.

In the present case the coefficients $C_{n}$ are given by those of the TSRC.
As stressed before, when $N\gtrapprox 12$, squeezing is lost and TSRC has
its sub-poissonian statistics changed to a super-poissonian one. Thus, in
the following, when considering values of\ $N$ it can be interesting to take
into account these two boundaries. Table I and II show the roots $\beta
_{k}^{\ast }=\left\vert \beta _{k}\right\vert e^{-i\varphi _{\beta _{k}}}$
of the characteristic polynomial in Eq.(\ref{raizes}) and the displacement
parameters $\alpha _{k}=\left\vert \alpha _{k}\right\vert e^{-i\varphi
_{\alpha _{k}}}$ for $N=5$, and $N=13$, respectively. The probability of
successfully engineering TSRC, $P_{\left\vert TSRC\right\rangle }$, diminish
with the number $N$, as is true for any state.

\begin{table}[tbp]
\caption{The roots $\protect\beta _{k}^{\ast }=|\protect\beta _{k}|e^{-i%
\protect\varphi _{\protect\beta _{k}}}$ of the characteristic polynomial and
the displacement parameters $\protect\alpha _{k}^{\ast }=|\protect\alpha %
_{k}|e^{-i\protect\varphi _{\protect\alpha _{k}}}$ for TSRC. Here, $N=5$, $%
T=0.878$. The probability of successfully engineering TSRC is 0.2$\%$}%
\centering$%
\begin{tabular}{||c||c||c||c||c||}
\hline\hline
N & $\left| \beta _{k}\right| $ & $\varphi _{\beta _{k}}$ & $\left| \alpha
_{k}\right| $ & $\varphi _{\alpha _{k}}$ \\ \hline\hline
1 & 1.465 & 3.141 & 0.883 & 0.297 \\ \hline\hline
2 & 1.080 & 2.307 & 0.647 & -2.316 \\ \hline\hline
3 & 1.080 & -2.307 & 1.083 & 1.570 \\ \hline\hline
4 & 2.190 & 1.283 & 2.465 & -2.005 \\ \hline\hline
5 & 2.190 & -1.283 & 3.688 & 1.570 \\ \hline\hline
6 &  &  & 2.190 & -1.283 \\ \hline\hline
\end{tabular}%
$%
\end{table}
%

\begin{table}[tbp]
\caption{The roots $\protect\beta _{k}^{\ast }=|\protect\beta _{k}|e^{-i%
\protect\varphi _{\protect\beta _{k}}}$ of the characteristic polynomial and
the displacement parameters $\protect\alpha _{k}^{\ast }=|\protect\alpha %
_{k}|e^{-i\protect\varphi _{\protect\alpha _{k}}}$ for TSRC. Here, $N=13$
and $T=0.951$. The probability of successfully engineering TSRC is $%
10^{-6}\% $}\centering$%
\begin{tabular}{||c||c||c||c||c||}
\hline\hline
N & $\left| \beta _{k}\right| $ & $\varphi _{\beta _{k}}$ & $\left| \alpha
_{k}\right| $ & $\varphi _{\alpha _{k}}$ \\ \hline\hline
1 & 3.169 & 3.141 & 1.149 & 0.102 \\ \hline\hline
2 & 2.651 & 2.465 & 1.089 & -2.156 \\ \hline\hline
3 & 2.651 & -2.465 & 1.908 & 1.570 \\ \hline\hline
4 & 1.529 & 3.141 & 1.055 & -1.885 \\ \hline\hline
5 & 0.701 & 3.141 & 0.526 & 3.141 \\ \hline\hline
6 & 2.414 & 1.731 & 1.608 & -1.702 \\ \hline\hline
7 & 2.414 & -1.731 & 3.353 & 1.570 \\ \hline\hline
8 & 3.498 & 1.420 & 4.374 & -1.725 \\ \hline\hline
9 & 3.498 & -1.420 & 5.381 & 1.570 \\ \hline\hline
10 & 1.466 & 0.921 & 3.796 & -1.648 \\ \hline\hline
11 & 1.466 & -0.921 & 2.009 & -1.648 \\ \hline\hline
12 & 2.772 & 0.574 & 2.747 & -2.065 \\ \hline\hline
13 & 2.772 & -0.574 & 2.865 & 1.570 \\ \hline\hline
14 &  &  & 2.772 & -0.574 \\ \hline\hline
\end{tabular}%
$%
\end{table}
For $N=5$, $P_{\left\vert TSRC\right\rangle }=0.2\%$. The beam-splitter
transmittance which optimizes this probability is $T=0.878$. For $N=13$, $%
P_{\left\vert TSRC\right\rangle }=10^{-6}\%$ and the optimized beam-splitter
transmittance is $T=0.951$.

\section{Fidelity of generation of TSRC}

In the prior sections we have assumed ideal detectors and beam-splitters.
Although very good beam-splitters are available by advanced technology
(losses by absorption lesser than $2\%$), the same is not true for
photo-detectors in the optical domain. Thus, we now take into account the
quantum efficiency $\eta $ at the photodetectors. For this purpose, we use
the Langevin operator technique as introduced in \cite{norton1} to obtain
the fidelity to get the TSRC.

Output operators accounting for the detection of a given field $\hat{\alpha}$
reaching the detectors are given by \cite{norton1}

\begin{equation}
\widehat{\alpha }_{out}=\sqrt{\eta }\widehat{\alpha }_{in}+\widehat{\mathsf{L%
}}_{\alpha },  \label{E9}
\end{equation}%
where $\eta $ stands for the efficiency of the detector and $\widehat{%
\mathsf{L}}_{\alpha }$, acting on the environment states, is the noise (or
Langevin) operator associated with losses into the detectors placed in the
path of modes $\widehat{\alpha }=a,b$. We assume that the detectors couple
neither different modes $a,b$ nor the Langevin operators $\widehat{\mathsf{L}%
}_{a}$,$\widehat{\mathsf{L}}_{b}.$ Thus, the following commutation relations
are readily obtained from Eq.(\ref{E9}):

\begin{eqnarray}
\left[ \widehat{\mathsf{L}}_{\alpha },\widehat{\mathsf{L}}_{\alpha
}^{\dagger }\right] &=&1-\eta ,  \label{E10a} \\
\left[ \widehat{\mathsf{L}}_{\alpha },\widehat{\mathsf{L}}_{\beta }^{\dagger
}\right] &=&0\text{,}  \label{E10b}
\end{eqnarray}%
which give rise to following ground-state expectation values for pairs of
Langevin operators:

\begin{eqnarray}
\left\langle \widehat{\mathsf{L}}_{\alpha }\widehat{\mathsf{L}}_{\alpha
}^{\dagger }\right\rangle &=&1-\eta ,  \label{E11a} \\
\left\langle \widehat{\mathsf{L}}_{\alpha }\widehat{\mathsf{L}}_{\beta
}^{\dagger }\right\rangle &=&0\text{.}  \label{E11b}
\end{eqnarray}%
These are useful relations specially for optical frequencies, when the state
of the environment can be very well approximated by the vacuum state, even
for room temperature.

Let us now apply the scheme of the Ref.\cite{dakna} to the present case. For
simplicity we will assume all detectors having high efficiency ($\eta
\gtrsim 0.9$). This assumption allows us to simplify the resulting
expression by neglecting terms of order higher than $(1-\eta )^{2}$. When we
do that, instead of the ideal $|TSRC\rangle $, we find the mixed state $%
|\Psi _{FE}\rangle $ describing the field plus environment, the latter being
due to losses coming from the nonunit efficiency detectors. The result is,
\begin{eqnarray}
\left\vert \Psi _{FE}\right\rangle &\sim &R^{N}D(\alpha _{N+1})\hat{a}^{+}T^{%
\hat{n}}D(\alpha _{N})\hat{a}^{+}T^{\hat{n}}  \notag \\
&\times &D(\alpha _{N-1})\ldots \hat{a}^{+}T^{\hat{n}}D(\alpha _{1})\text{ }%
\left\vert 0\right\rangle \widehat{\mathsf{L}}_{0}^{\mathsf{+}}  \notag \\
&+&R^{N-1}D(\alpha _{N+1})\hat{a}^{+}T^{\hat{n}}D(\alpha _{N})\hat{a}^{+}T^{%
\hat{n}}  \notag \\
&\times &D(\alpha _{N-1})\ldots \widehat{\mathsf{L}}_{1}^{+}T^{\hat{n}%
}D(\alpha _{1})\left\vert 0\right\rangle  \notag \\
&+&R^{N-1}D(\alpha _{N+1})\hat{a}^{+}T^{\hat{n}}D(\alpha _{N})\widehat{%
\mathsf{L}}_{N-1}^{+}T^{\hat{n}}  \notag \\
&\times &D(\alpha _{N-1})\ldots \hat{a}^{+}T^{\hat{n}}D(\alpha _{1})\text{ }%
\left\vert 0\right\rangle  \notag \\
&+&R^{N-1}D(\alpha _{N+1})\widehat{\mathsf{L}}_{N}^{+}T^{\hat{n}}D(\alpha
_{N})\hat{a}^{+}T^{\hat{n}}  \notag \\
&\times &D(\alpha _{N-1})\ldots \hat{a}^{+}T^{\hat{n}}D(\alpha _{1})\text{ }%
\left\vert 0\right\rangle ,  \label{damped}
\end{eqnarray}%
where, for brevity, we have omitted the kets corresponding to the
environment. Here $R$ is the reflectance of the beam splitter, $\widehat{%
\mathsf{L}}_{0}^{\mathsf{+}}=\mathbf{1}$ is the identity operator, and $%
\widehat{\mathsf{L}}_{k}$, $k=1,2..N$ stands for losses in the first, second
$\ldots $ N-$th$ detector. Although the $\widehat{\mathsf{L}}_{k}\prime s$
commute with any system operator, we have maintained the order above to keep
clear the set of possibilities for photo absorption: the first term, which
includes $\widehat{\mathsf{L}}_{0}^{\mathsf{+}}=\mathbf{1}$, indicates the
probability for nonabsorption; the second term, which include $\widehat{%
\mathsf{L}}_{1}^{+}$, indicates the probability for absorption in the first
detector; and so on. Note that in case of absorption at the k-$th$ detector,
the annihilation operator $a$ is replaced by the $\widehat{\mathsf{L}}%
_{k}^{+}$ creation Langevin operator. Other possibilities such as absorption
in more than one detector lead to a probability of order lesser than $%
(1-\eta )^{2}$, which are neglected.

Next, we have to compute the fidelity \cite{nota}, $F=\left\Vert
\left\langle \Psi \right. \left\vert \Psi _{FE}\right\rangle \right\Vert
^{2} $, where $\left\vert \Psi \right\rangle $ is the ideal state given by
Eq.(\ref{experimental}), here corresponding to our CCS characterized by the
parameters shown in Table 1, and $\left\vert \Psi _{FE}\right\rangle $ is
the (mixed) state given in the Eq.(\ref{damped}). Assuming $\eta =$ $0.950$
and $0.900$ we find, for $N=5$ ($13$), $F=$ $0.997$ ($0.994$) and $0.996$ ($%
0.991$), respectively. These high values of fidelities show that
efficiencies around $0.9$ lead to states whose degradation due to losses is
not so dramatic.

\section{Comments and conclusion}

In this paper we introduce a new state of the quantized electromagnetic
field, the Truncated State with Random Coefficients (TSRC). To characterize
the TSRC, we have studied several of its statistical properties as well as
the behaviour of these statistical properties when the dimension $N$ of
Hilbert space is increased. By studying the dependence of the $Q$-parameter
and of the second order correlation function with increasing $N$, we found
that there exists a transition from sub-poissonian statistics to
super-poissonian statistics when $N$ is relatively small ($N\sim 12$).
Interesting, we have also found that for $N\gtrsim 12$ squeezing is rarely
observed. Other statistical properties such as Shannon entropy, quadrature
operator and their variances were investigated. We think \ TSRC \ can be
used to characterize sequence of numbers generated deterministically. It
would be interesting, for example, to compare TSRC properties with the
properties of states whose coefficients stem from a sequence generated by
iteration of the logistic equation.

\section{Acknowledgments}

We thank B. Baseia and A. T. Avelar for the carefully reading of this
manuscript and valuable suggestions, the CAPES (WBC) and the CNPq (NGA),
Brazilian agencies, for the partial supports of this work.


\begin{thebibliography}{99}
\bibitem{bennett93} C. H. Bennett et. al. Phys. Rev. Lett. \textbf{70 }%
(1993) 1895.

\bibitem{Kane98} B. E. Kane, Nature, \textbf{393}, 143(1998), and references
therein.

\bibitem{Pellizzari97} T. Pellizzari, Phys. Rev. Lett. \textbf{79},
5242(1997), and references therein.

\bibitem{Gisin02} N. Gisin, G. Ribordy, W. Titel and H. Zbinden, Rev. Mod.
Phys. \textbf{74}, 145(2002).

\bibitem{Bjork01} see,e.g., G. Bj\"{o}rk and L.L. Sanchez-Soto, Phys. Rev.
Lett., \textbf{86}, 4516(2001) ; M. M\"{u}tzel \ et al., Phys. Rev. Lett.
\textbf{88}, 083601(2002), \ and refs.therein.

\bibitem{Zurek91} W.H. Zurek, Phys. Today, \textbf{44}, 36 (1991); C.C.
Gerry and P.L. Knight, Am. J. Phys., \textbf{65}, 964 (1997); B.T.H. Varcoe
et al., Nature, \textbf{403}, 743 (2000).

\bibitem{harochegato} J. M. Raimond, M. Brune, and S. Haroche, Phys. Rev.
Lett. \textbf{79}, 1964 (1996); S. Osnaghi, P. Bertet, A. Auffeves, P.
Maioli, M. Brune, J. M. Raimond, and S. Haroche, Phys. Rev. Lett. \textbf{87}%
, 37902 (2001).

\bibitem{brune} M. Brune et. al., Phys. Rev. Lett. \textbf{77} (1996) 4887.

\bibitem{bennett2} C. H. Bennett, D. P. Vicenzo, Nature \textbf{404} (2000)
247; A. K. Ekert, Phys. Rev. Lett. \textbf{67} (1991) 661.

\bibitem{narozhny} N. B. Narozhny, J. J. Sanchez-Mondragon, J. H. Eberly,
Phys. Rev. A \textbf{23} (1981) 236; G. Rempe, H. Walther, N. Klein, Phys.
Rev. Lett. \textbf{58} (1987) 353.

\bibitem{zoller} J. F. Poyatos, J. I. Cirac, and P. Zoller, Phys. Rev. Lett.
\textbf{77} (1996) 4728.

\bibitem{barnett} S. M. Barnett, D. T. Pegg, Phys. Rev. Lett. \textbf{76}
(1996) 4148; G. Bjork, L. L. Sanchez-Soto, J. Soderholm, Phys. Rev. Lett.
\textbf{86} (2001) 4516.

\bibitem{serra} R. M. Serra, N. G. de Almeida, C. J. Villas-B\^{o}as, and M.
H. Y. Moussa, Phys. Rev A. \textbf{62} (2000) 43810.

\bibitem{wesley} W. B. Cardoso and N. G. de Almeida, "\textit{Truncated
states obtained by iteration}", In preparation.

\bibitem{serra2} R. M. Serra, P.B. Ramos, N. G. de Almeida, W. D. Jos\'{e},
and M. H. Y. Moussa, Phys. Rev. A \textbf{63} (2001) 053803.

\bibitem{vogel} K. Vogel, V. M. Akulin, and W. P. Schleich, Phys. Rev. Lett
\textbf{71} (1993) 1816; M. H. Y. Moussa and B. Baseia Phys. Lett. A 238
(1998) 223.

\bibitem{dakna} M. Dakna, J. Clausen, L. Kn\"{o}ll and D.-G. Welsch, Phys.
Rev. A \textbf{59}, 1658(1999).

\bibitem{Marhic} M. E. Marhic, P. Kumar, Optics Commun. \textbf{76} (1990)
143.

\bibitem{walls} D. F. Walls, G. J. Milburn, \textit{Quantum Optics},
Springer-Verlag, (Berlin, 1994).

\bibitem{norton1} C. J. Villas-Boas, N. G. de Almeida and M. H. Y. Moussa,
Phys. Rev. A \textbf{60} (1999) 2759.

\bibitem{nota} The expression $F=\| \langle TSRC | \Psi _{FE}\rangle \|^{2}$
stands for usual abbreviation in the literature. Actually, this is
equivalent to $\langle\Psi_{TSRC}|Tr_{E}\hat{\varrho}_{FE}|\Psi_{TSRC}%
\rangle $ where $\hat{\varrho}_{FE}=|\Psi_{FE}\rangle\langle\Psi_{FE}|$ and $%
Tr_{E}\hat{\varrho}_{FE}$ is the (\textit{mixed}) smeared state, mentioned
before.
\end{thebibliography}
\end{document}